\documentclass[10pt,journal]{IEEEtran}

\usepackage{graphicx} 
\usepackage{gensymb} 
\usepackage{ragged2e} 
\usepackage{amsmath} 

\ifCLASSOPTIONcompsoc
  \usepackage[nocompress]{cite}
\else
  \usepackage{cite}
\fi

\begin{document}

\title{Implications of Integrated CPU-GPU Processors on Thermal and Power Management Techniques}

\author{Kapil~Dev,
        Indrani~Paul,
        Wei~Huang,
        Yasuko~Eckert,
        Wayne~Burleson,
        and~Sherief~Reda

\thanks{This work was done at Brown University, and in collaboration with Advanced Micro Devices, Pvt. Ltd. } 
\thanks{K. Dev and S. Reda:  School of Engineering, Brown University, Providence, RI. Contact email: kapil\_dev@alumni.brown.edu.}
\thanks{I. Paul, W. Huang, Y. Eckert, and W. Burleson: Advanced Micro Devices, Pvt. Ltd.}
\vspace{-0.1in}
}

\markboth{}%
{}

\maketitle

\begin{abstract}
Heterogeneous processors with architecturally different cores (CPU and GPU) integrated on the same die lead to new challenges and opportunities for thermal and power management techniques because of shared thermal/power
budgets between these cores. In this paper, we show that new parallel programming paradigms (e.g., OpenCL) for
CPU-GPU processors create a tighter coupling between the workload, the thermal/power management unit and the operating system. Using detailed thermal and power maps of the die from infrared imaging, we demonstrate that in contrast to traditional multi-core CPUs,  heterogeneous processors exhibit higher coupled behavior for dynamic voltage and frequency scaling and workload scheduling, in terms of their effect on performance, power, and temperature. Further, we show that by taking the differences in core architectures and relative proximity of different computing cores on the die into consideration, better scheduling schemes could be implemented to reduce both the power density and peak temperature of the die. The findings presented in the paper can be used to improve thermal and power efficiency of heterogeneous CPU-GPU processors.
\end{abstract}

\begin{IEEEkeywords}
Heterogenous CPU-GPU processors, infrared imaging, OpenCL workloads, thermal and power management, .
\vspace{-0.05in}
\end{IEEEkeywords}

\vspace{-0.2in}
\ifCLASSOPTIONcompsoc
\IEEEraisesectionheading{\section{Introduction}\label{sec:introduction}}
\else
\section{Introduction}
\label{sec:introduction}
\fi

\noindent Typically, CPU and GPU devices are optimized for different application domains. However, heterogeneous processors with both CPU and GPU devices integrated on the same die provide good power efficiency (performance/Watt) for both single-threaded, control-divergence dominated and data parallel workloads~\cite{Daga_saahpc2011, RezaCuster17, TannCodes16}. Therefore, heterogeneous CPU-GPU processors are becoming mainstream these days. Further, the new programming paradigms (e.g., OpenCL) for these processors allow arbitrary work-distribution between CPU and GPU devices, where the programmer controls the distribution at the application development time~\cite{OpenCL}. Due to the shared nature of thermal and power resources and due to application dependent work distribution between two devices, there are new challenges and opportunities to optimize performance and power efficiency of the CPU-GPU processors~\cite{devJOLPE17, PaulISCA13, DevESTIMedia16, MajumdarIISWC15, PaulSC13, DevIISWC16}.

Modern processors have two main knobs of thermal and power management: dynamic voltage and frequency scaling (DVFS), and scheduling of workloads on different compute units of the chip~\cite{Eyerman_TACO2011, Donald_isca2006}. DVFS is used to trade performance for keeping temperature and power below their safe limits; similarly, thermal-aware scheduling helps in distributing thermal hot spots across the die. 
In a traditional multi-core CPU, all cores have the same micro-architecture; therefore, at a fixed DVFS setting scheduling has little or negligible effect on the performance and power of a workload, especially for a single threaded workload. On the other hand, as we demonstrate in this paper, DVFS and spatial scheduling based power management decisions are highly intertwined in terms of performance and power efficiency tradeoffs on a heterogeneous CPU-GPU processor. 

To understand the impact of thermal and power management techniques, one needs to have a setup that can provide detailed thermal and power maps of the die while it is running workloads. Recent works have proposed using infrared (IR) camera based thermal imaging techniques for the detailed post-silicon thermal and power mapping of integrated circuits~\cite{Cochran_islped2010, redaBPIsensors, PaekTCAD13,Huang09,Martinez_isca2007,DevISLPED13,Kursun08}. While Cochran \emph{et al.} proposed techniques to obtain detailed power maps from the IR-based steady-state thermal images of FPGA~\cite{Cochran_islped2010}, Paek \emph{et al.} proposed Markov Random Field based framework for transient and steady-state analysis of FPGAs\cite{PaekTCAD13}. These works implemented relatively low power circuits, so they did not require any special heatsink system for cooling the FPGA.  On the other hand, thermal imaging of relatively high-power processors requires a special heatsink setup which is transparent to IR emissions. While Huang \emph{et al.} model the differences between thermal profiles of a fluid-based and fan-based heatsink setups~\cite{Huang09}, Martinez \emph{et al.} and Dev \emph{et al.} used IR-transparent oil-based heatsink setups to perform detailed thermal and power mapping on general-purpose processors~\cite{Martinez_isca2007,DevISLPED13}. Further, Kursun \emph{et al.} used IR imaging for studying the effect of process variations on the thermal profiles of a multi-core test-chip~\cite{Kursun08}. These prior works focused on the CPU-only processors, while our focus in this work is on CPU-GPU heterogeneous processors, which exhibit different thermal and power profiles than the CPU-only processors because they have two architecturally different devices integrated on the same die. To the best of our knowledge, our work is the first to perform detailed thermal and power mapping of a heterogeneous CPU-GPU processor through experiments. 

Our experiments use a wide array of physical measurements on a real heterogeneous processor, which enable us to capture many of the intricacies of real hardware. In particular, our measurements include high-resolution thermal maps of the processor's die using infrared imaging, and power maps computed by inverting the thermal maps with high accuracy. 
In addition to confirming simulated behavior, our experiments highlight multiple implications of CPU-GPU processors on thermal and power management techniques. The main contributions of this work are as follows. 

 \begin{figure*}[t!]
\begin{center}
\scalebox{0.75}{
\includegraphics{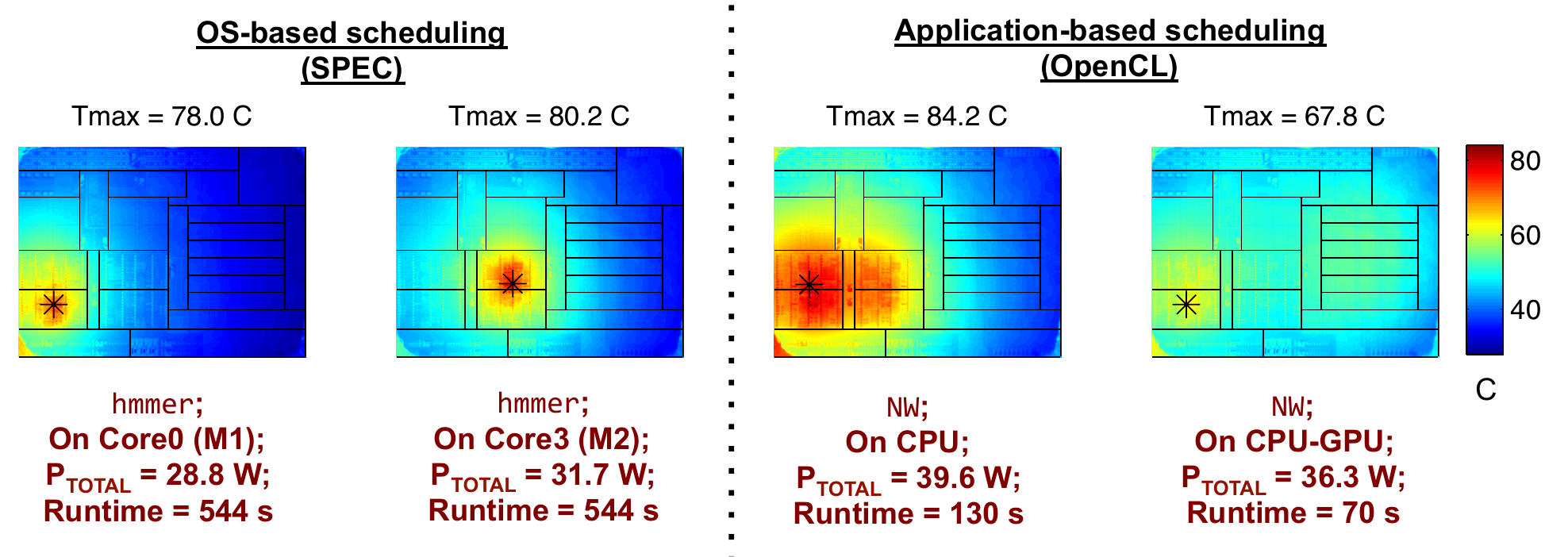}}
\vspace{-0.15in}
\caption{Scheduling techniques: OS-based scheduling of the SPEC\textsuperscript{\textregistered} CINT 2006 benchmark (\texttt{456.hmmer}) and application-based scheduling of an OpenCL benchmark (\texttt{NW}).}
\vspace{-0.25in}
\label{fig:os_based_vs_app_based}
\end{center}
\end{figure*}

\begin{itemize}
\item We use infrared imaging to characterize the effects of task scheduling and DVFS on a heterogeneous processor. In particular, we provide detailed thermal and power breakdown of a real CPU-GPU chip as a function of hardware module and workload characteristics.
\item We show that compared to traditional operating system-based scheduling, the application-based scheduling using OpenCL Runtime can achieve more thermal and power efficient scheduling by taking advantage of the heterogeneity in a CPU-GPU processor.
\item We study interactions between workload characteristics, scheduling decisions, and DVFS settings for OpenCL workloads. 
We observe that the effects of DVFS and scheduling on performance, power, and temperature for OpenCL workloads are  intertwined. Therefore, DVFS and scheduling must be considered simultaneously to achieve the optimal runtime and energy on CPU-GPU processors.
\item We show that CPU and GPU devices have different power densities and thermal profiles, which could have multiple implications on the thermal and power management solutions for such processors.
\item We demonstrate that a leakage power-aware scheduling on CPU cores for GPU workloads could reduce both the peak temperature and total power of the chip.
\end{itemize}
 
The rest of the paper is organized as follows. Section~\ref{sec:motivation} provides the motivation behind this work by highlighting the differences in thermal profiles of CPU-only and CPU-GPU processors. We provide details of the experimental setup, modeling and simulation setup, and benchmarks used in this work  in Section~\ref{sec:exp_setup}. Section~\ref{sec:results} presents the experimental results characterizing the thermal and power behavior of a real CPU-GPU processor.
Finally, Section~\ref{sec:conclusions} concludes and provides future directions for the work.


\vspace{-0.05in}
\section{Motivation}
\label{sec:motivation}

CPU-GPU processors have different thermal and power management mechanisms than traditional multi-core CPUs because they have two architecturally dissimilar devices integrated on the same die. For example, in a traditional multi-core CPU, workload scheduling is done at operating system (OS) level using system level commands like \textit{taskset}~\cite{LinuxTaskset}. However, in CPU-GPU processors, the OpenCL framework provides application-programming interfaces (APIs) to control the compute devices; it allows the programmer to distribute the work among different devices. The inherently more complex scheduling factors, combined with architecturally different devices, leads to different thermal and power profiles in CPU-GPU processor than the traditional CPU-only processor.

\begin{figure}[t!]
\begin{center}
\scalebox{1}{
\includegraphics[width=0.3\textwidth]{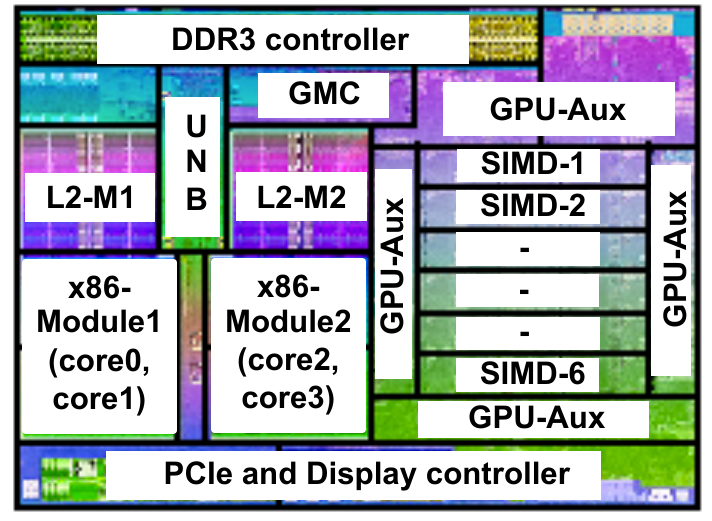}}
\vspace{-0.05in}
\caption{Floorplan of the AMD A10-5700 APU.}
\vspace{-0.25in}
\label{fig:floorplan}
\end{center}
\end{figure}

Figure~\ref{fig:os_based_vs_app_based} depicts the outcome of the following two scheduling choices: termed as purely OS-based and application-based scheduling here, with the help of thermal maps for a SPEC CPU benchmark (\texttt{hmmer}) and an OpenCL workload (\texttt{NW})~\cite{CheIISWC09}. The floorplan of the processor is also overlaid on these thermal maps. The die-shot with floorplan information of the processor is shown in Figure~\ref{fig:floorplan}. From the Figure~\ref{fig:os_based_vs_app_based}, we observe that with the help of OS, the  \texttt{hmmer} benchmark could be launched on one of the CPU cores as shown in the thermal maps in first two columns. As expected, the thermal hotspots are primarily located in the active cores (e.g., on core0 and core3) in these maps. On the other hand, with the help of OpenCL runtime, the application-based scheduler could launch the kernels of \texttt{NW} workload  on both CPU and GPU devices. The thermal maps for such scheduling are shown in the 3$^{rd}$ and 4$^{th}$ columns in Figure~\ref{fig:os_based_vs_app_based}. We notice that the thermal profiles of OpenCL workload are different. In particular, when the OpenCL kernels are launched on CPU, they use all four cores of the CPU; so, the thermal hotspots in this case are spread over all cores (as shown in the thermal map in $3^{rd}$ column). Further, when the kernels are launched on GPU, the hotspots are spread over both CPU and GPU, with CPU being hotter than GPU due to its higher power density. This is also because, in the \texttt{NW} benchmark, when the GPU is running a particular iteration of kernels, the CPU is preparing work for the future iterations, keeping both CPU and GPU devices active simultaneously.

Further, from the Figure~\ref{fig:os_based_vs_app_based}, we observe that the application-based scheduling on a CPU-GPU processor leads to larger range of the peak temperature (84.2~$\celsius$ and 67.8~$\celsius$) and total runtime (130~s and 70~s) by running workloads on CPU and GPU devices than the CPU-only scheduling from the OS. Hence, we observe that although both the OS and application based scheduling affect the thermal hot spot locations, the application based scheduling inherently takes advantage of the heterogeneity in a CPU-GPU system and achieves a more thermal friendly and power efficient scheduling as well as achieving better performance. 

In summary, scheduling on a CPU-GPU processor creates tight interaction between the application and the power management unit.  Thus, a scheduling scheme could be implemented on such processors wherein the thermal/power management unit, OS and application can make the scheduling decisions in collaboration. In this paper, we highlight multiple such implications of integrating CPU-GPU devices on a single die through experiments on a real processor.


\vspace{-0.05in}
\section{Background and Setup Details}
\label{sec:exp_setup}

Understanding the impact of thermal and power management techniques requires a setup that can be used to obtain detailed thermal and power maps of the die while it is running each  workload. 
To this end, we developed an infrared imaging setup to capture the thermal images of a processor in real time. The detailed power maps are obtained by inverting the thermal images using the thermal-to-power model of the die with the cooling system. 
Below, we describe the experimental and simulation setup, along with the benchmarks, used to obtain the thermal and power maps of the die in this work.

\vspace{-0.1in}
\subsection{CPU-GPU Processor and Motherboard}
Our experimental system consists of a motherboard with FM2+ socket fitted with an AMD A10-5700 Accelerated Processing Unit (APU) and 8 GB DRAM as a pair of 4 GB DDR3 DIMMs. The floorplan of the APU is shown in Figure~\ref{fig:floorplan}~\cite{trinity_hc2012}. The APU has two x86 modules, each containing a 2$\times$2MB L2 cache and two CPU cores. The APU has an integrated GPU with 6 single instruction multiple data (SIMD) compute units. The IP blocks surrounding the SIMD units are denoted as GPU auxiliary units because they are active when the SIMD units are active. The area between L2 caches is occupied by unified north bridge (UNB), which acts as an interface between L2 caches and DDR3 controller. The GPU has an additional memory controller, called graphics memory controller (GMC), which is optimized for the graphics related memory requests. In this paper we consider two frequency settings for CPU: 1.4 GHz and 3.0 GHz. The APU and the motherboard used in this study has no BIOS-level support for frequency scaling of the GPU device (only DFS without voltage scaling) and hence, limits the usefulness for the power management. So, we fix the GPU frequency to 800 MHz in our experiments. The latest AMD APUs may support full DVFS on the integrated GPU, thus providing more potential for power, energy, and thermal control than shown in this work.

\vspace{-0.1in}
\subsection{IR-based Thermal Imaging and Custom Heatsink Setup} 
We have developed an IR-imaging based lab setup that allows us to perform detailed thermal and power mapping of CPU-GPU processors. To capture the thermal emissions from the back-side of the die, we use an FLIR SC5600 infrared camera with a mid-wave spectral range of 2.5 -- 5.1 $\mu$m.  The camera's cryogenic cooled InSb detectors have a sensitivity of about 15 mK noise equivalent temperature difference (NETD) and is capable of operating at 380 Hz at 640$\times$512 pixels resolution. We used a similar setup to perform detailed power mapping of a quad-core CPU processor~\cite{DevISLPED13}. While the basic setup and power mapping framework remains the same when we change the processor and/or the motherboard, the IR-transparent heatsink has to be changed whenever the physical dimensions of the CPU socket and heatsink assembly changes. The A10-5700 APU uses an FM2+ socket,  so, we built a new IR-transparent heatsink for the CPU-GPU processor.

 \begin{figure}[t!]
\begin{center}
\vspace{-0.1in}
\scalebox{0.6}{
\includegraphics{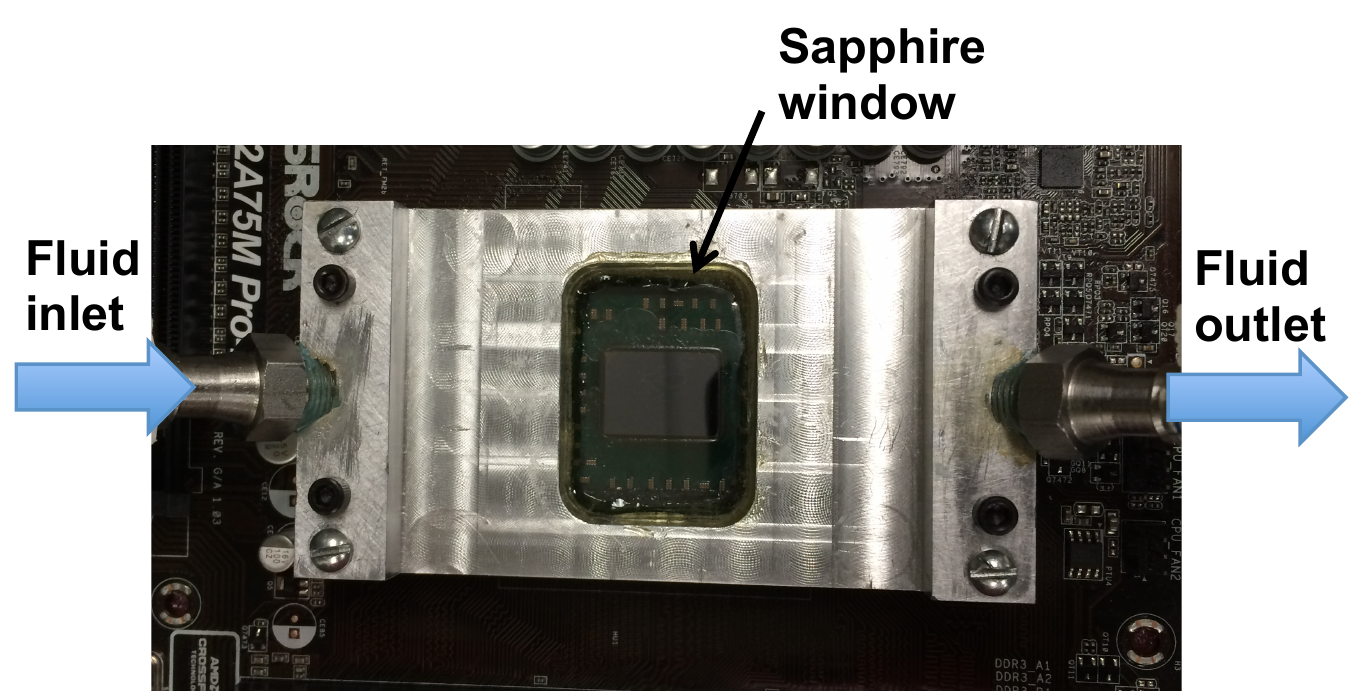}}
\caption{Infrared-transparent oil-based heat removal system.}
\vspace{-0.25in}
\label{fig:oilSystem}
\end{center}
\end{figure}

The custom heatsink system for the FM2+ socket is shown in Figure~\ref{fig:oilSystem}. It has a rectangular channel of height 1 mm through which an infrared-transparent mineral oil is flowing from the inlet valve to the outlet valve. Two infrared-transparent sapphire windows (one at the top and other one at bottom of channel) are assembled in the system in such a way that they allow midwave infrared waves to pass through part of the channel. In addition to being infrared transparent, the bottom window also spreads the heat generated in small processor die over a larger area, which improves heat removal capacity of the system. Further, the IR-transparent oil itself is used as the thermal interface material (TIM) between the bottom window and the processor die. The TIM fills any potential micro air-gaps between the two surfaces, thereby improving the overall heat transfer at the window-die interface.

\newcommand{\tbold}{\mathbf{t}}
\newcommand{\ubold}{\mathbf{u}}
\newcommand{\Rbold}{\mathbf{R}}
\newcommand{\Kbold}{\mathbf{K}}
\newcommand{\Cbold}{\mathbf{C}}
\newcommand{\pbold}{\mathbf{p}}
\newcommand{\Pbold}{\mathbf{P}}
\newcommand{\fbold}{\mathbf{f}}
\newcommand{\Ubold}{\mathbf{U}}

\newcommand{\vbold}{\mathbf{v}}
\newcommand{\Ibold}{\mathbf{I}}
\newcommand{\Fbold}{\mathbf{F}}

\vspace{-0.1in}
\subsection{Modeling Relationship Between Temperature and Power}  
Our goal is to model the relationship between power and temperature on a processor for the custom heatsink system. In particular, if $\mathbf{t}$ is a vector that denotes the steady state or averaged  thermal map of the processor in response to some power map denoted by $\mathbf{p}$, then our goal is to model the relationship between $\mathbf{p}$ and $\mathbf{t}$. Here, each element of $\mathbf{t}$ and $\mathbf{p}$ vectors denotes temperature and power of a particular block of the chip, respectively. To this end, we modeled the custom heatsink along with the CPU-GPU processor with the COMSOL Multiphysics tool, which is a widely used software to solve multiple coupled physical phenomena~\cite{comsol}. COMSOL has a finite element analysis (FEM) based solver as its core computational-engine.

\begin{figure}[t!]
\begin{center}
\scalebox{0.5}{
\includegraphics{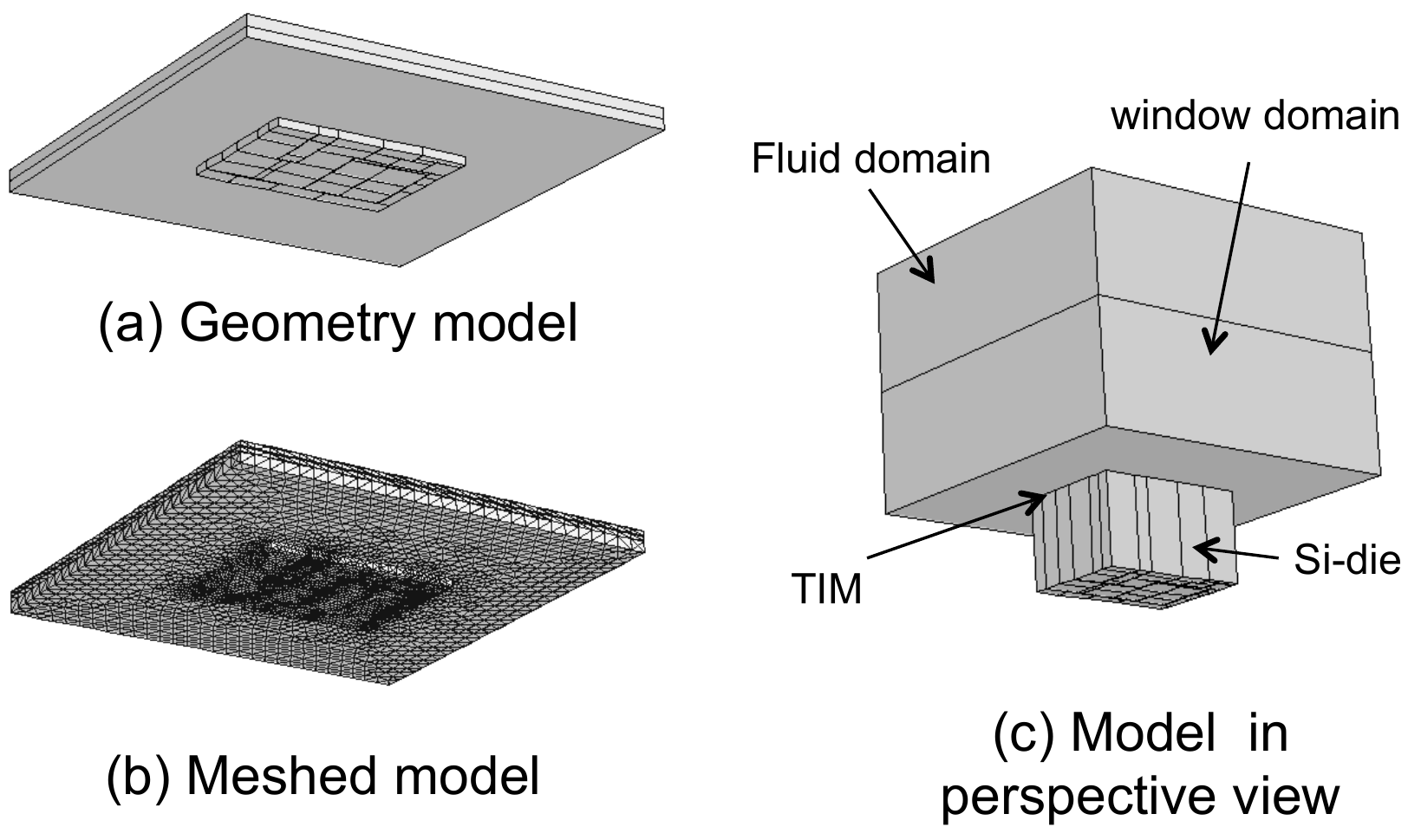}}
\vspace{-0.1in}
\caption{Model for the oil-based system: (a) model-geometry with actual aspect-ratio; (b) model-geometry in perspective view; (c) meshed model.}
\vspace{-0.25in}
\label{fig:oilHScomsol}
\end{center}
\end{figure}

The geometry of the simulated model is shown in Figure \ref{fig:oilHScomsol}.a; Figure \ref{fig:oilHScomsol}.b shows the meshed model used in the FEM simulation; Figure \ref{fig:oilHScomsol}.c shows a perspective view (not necessarily the same scale). The model has following domains: the processor's die, divided into a number of blocks as dictated by the floorplan, a 2 $\mu$m thermal interface material, an infrared-transparent sapphire-window and fluid domain. 
 The properties of different materials used in our simulation model are reported in Table \ref{table:materials}. 

\begin{table}[b!]
  \vspace{-0.15in}
\footnotesize
  \centering
    \caption{Material properties.  $\rho$: density in $kg/m^3$, $k$: thermal conductivity in $W/(mK)$, $C_p$: the specific heat capacity at constant pressure in $J/(kgK)$, $\mu$; dynamic viscosity in $Pas$, for fluids.}
  \begin{tabular}{l|c|c|c|c}
    Material & \textbf{$\rho$}  & \textbf{$k$}  & \textbf{$C_p$}   & \textbf{$\mu$} \\
    \hline
    \hline
    Mineral oil & 838 & 0.138 & 1670 & 14.246e-3 \\
        \hline
    Silicon & 2330 & 148 & 703 & - \\
    \hline
    Sapphire & 4050 & 35 & 761 & - \\
    \hline
  \end{tabular}
  \vspace{-0.1in}
  \label{table:materials}
\end{table}

Essentially, we simulate two types of physics: {\it fluid flow} and {\it conjugate heat transfer}, simultaneously to obtain the temperature profile for a given power dissipation profile of the processor. For the flow simulation, we use experimentally measured fluid flow-rate (1.4~gpm), inlet temperature (12.1~$\celsius$), and pressure (28~psi) as boundary conditions. We use Proteus Fluid Vision flow meter to measure the flow properties.
Internally, the FEM tool solves Navier-Stokes conservation of momentum and mass equations to simulate the flow. Second, for the heat transfer simulation, both in fluid and solid domains, the FEM tool solves the following heat-transfer equation in steady-state.

 \vspace{-0.1in}
\begin{equation}
\label{eqn:htSolidFluid}
\rho C_p \vbold . \nabla T = \nabla . \left( k \nabla T \right) + Q,
\end{equation}

\noindent where, $T$ is the temperature in Kelvin, $\vbold$ is the velocity field, and $Q$ denotes the heat sources in W/m$^3$.  For heat-transfer physics, we assume that all external walls of the system exchange heat with ambience through natural convection process; the typical heat-transfer coefficient ($h$) for the natural heat convection is 5 W/(m$^2$.K)~\cite{comsol}. In the simulation model, we use the silicon die of 750 $\mu m$ thick where the power dissipation happens at the bottom of the  silicon die. Assuming temperature independent material properties and fixed fluid velocity, equation~\ref{eqn:htSolidFluid} can be written in the following discrete linear form:

 \vspace{-0.15in}
\begin{equation}
\label{eqn:RpT}
\mathbf{R} \mathbf{p} =  \mathbf{t},
\end{equation}

\noindent
where, $\mathbf{R}$ is a {\it linear matrix operator} which encapsulates the relationship between the thermal and power profile of the die, $\mathbf{p}$ is the power map vector, where the power of each block, $p_i$, is represented by an element in $\mathbf{p}$, and $\mathbf{t}$ is the resultant temperature map of the die. The values of the matrix $\Rbold$ are learned through the FEM simulations of the setup, where we apply unit power pulses, one at a time, at each block location in simulation, and compute the thermal profile at the die-surface for each case. The thermal profile resulting from activating block $i$ corresponds to the $i^{th}$ column of $\mathbf{R}$. After simulating all blocks, we have the model matrix ($\Rbold$) complete. Similar approach has been used by the previous work to obtain the model matrix~\cite{Shakouri06a,Hamann07}. 

\noindent
{\bf Model Validation:}
To verify the accuracy of modeled matrix $\mathbf{R}$, we use a custom CPU-intensive micro-benchmark. First, we run the custom application on all four cores at the highest frequency and capture the steady-state thermal image of the die and measure the total power of the processor. Then, we change the frequency of just one core to ensure that the switching activity profile changes only in one core. We again capture a steady-state thermal image of the processor and measure total power. The difference in power map (say $ \delta \mathbf{p}$) is attributed to the core whose frequency is changed. Thus, we  compare the thermal simulation results of $\mathbf{R} \delta \mathbf{p}$ against the measured thermal image difference to verify the accuracy of the modeled $\mathbf{R}$ matrix.

\subsection{Thermal to Power Mapping} 
Reconstructing the underlying power map of the CPU-GPU processor from the measured thermal images is an inverse problem. Cochran et al.~\cite{Cochran_islped2010}, Martinez et al.~\cite{Martinez_isca2007}, and Qi et al.~\cite{Qi_iccd2010} describe techniques to solve the inverse problem of temperature to power mapping. Using similar techniques, in this work, we solve the following constrained least-square error minimization problem to reconstruct the power map ($\pbold$) of the die from its measured thermal map ($\tbold$).

 \vspace{-0.15in}
\begin{eqnarray}
\label{eq:PtotalTrain}
\mathbf{p}^* = & \arg_\pbold \min \Vert \Rbold \pbold - \tbold \Vert_2 \\
& \mbox{s. t.} \quad\ p_i \ge 0  \quad\quad\quad\quad\quad\quad\nonumber
\end{eqnarray} 

\noindent where, $\pbold^*$ is the reconstructed power-vector, $p_i$ denotes the power in the $i^{th}$ block of the die. By solving the above optimization problem, we obtain the total power of each block for the die. We use MATLAB inbuilt \emph{lsqlin} function to solve the above constrained least-square problem. Using, $p_i > 0$ constraint helps in ensuring that reconstructed power for all blocks is always positive. We also measure the total power of the processor by intercepting the 12 V supply lines going to the processor and measure the current through a shunt resistor connected to an external Agilent 34410A digital multimeter. The measured power is used to validate the accuracy of the power mapping process. In particular, we validated the reconstructed power values against the measured power and found that they have an average absolute inaccuracy of 3.01\%, which is sufficient for our purposes.

 \begin{figure*}[t!]
\begin{center}
\scalebox{0.7}{
\includegraphics{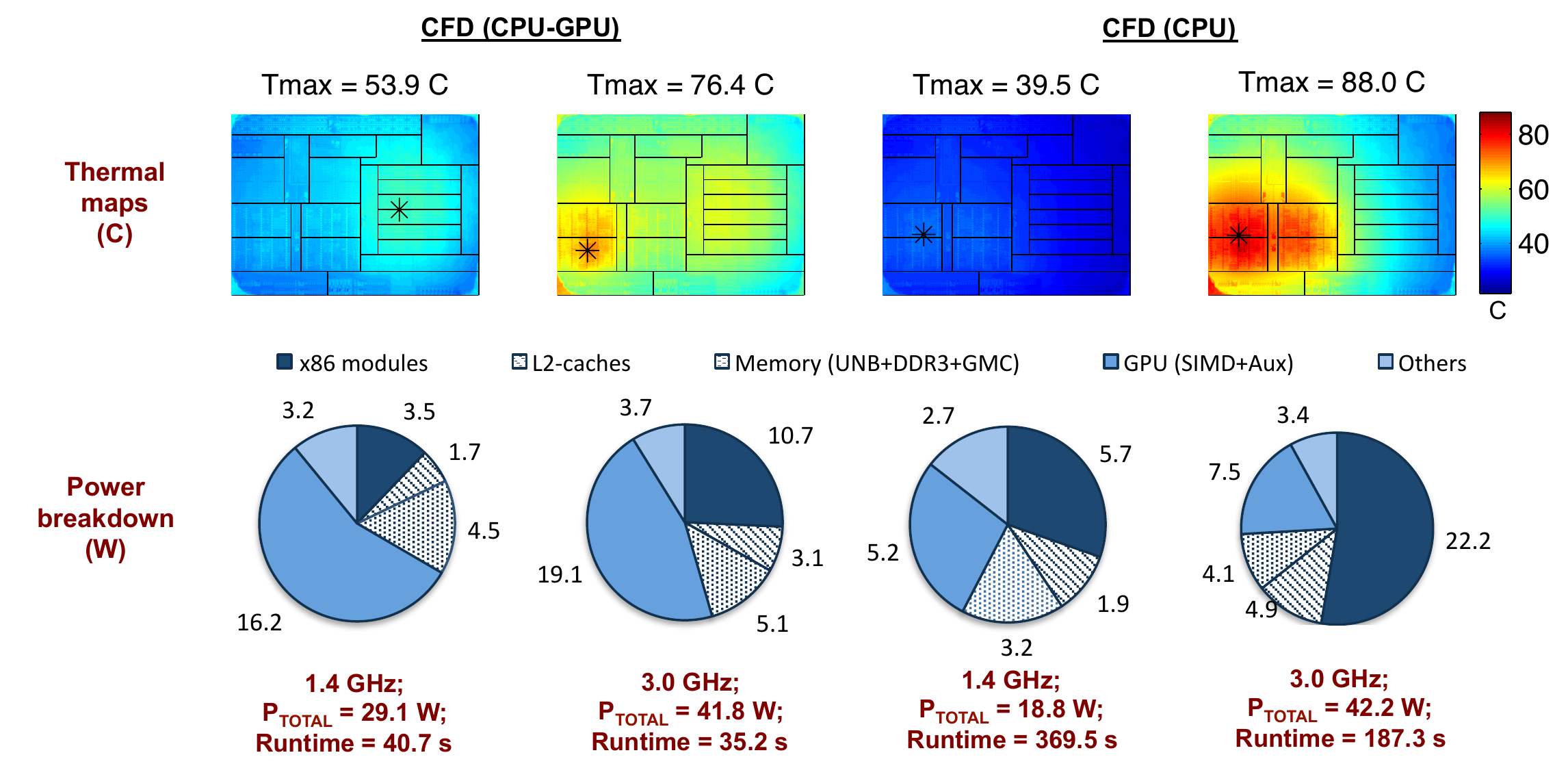}}
\vspace{-0.1in}
\caption{Thermal and power maps showing the interplay between DVFS and scheduling for \texttt{CFD} benchmark. The peak temperature, power and runtime are significantly different for different DVFS and scheduling choices.}
\vspace{-0.25in}
\label{fig:cfd_TPmaps}
\end{center}
\end{figure*}

\vspace{-0.1in}
\subsection{Benchmarks}  
In this work, we use the following workloads to cover a wide range of characteristics. First, to fully control the workload distribution between CPU and GPU devices, we wrote a simple custom micro-kernel (\texttt{$\mu$Kern}) in OpenCL that multiplies two vectors of arbitrary size for a given number of times. We use multiple iterations inside the kernel so that the die reaches a stable thermal state, which improves the reproducibility of thermal/power results. The micro-kernel is a \emph{homogeneous} type of workload because once it is launched on the GPU, the CPU is completely idle and vice-versa.
  
As a representation of real-life CPU-GPU workloads, we selected six OpenCL workloads from publicly available Rodinia benchmark suite~\cite{CheIISWC09}.
 In particular, we selected (\texttt{CFD}) solver from computational fluid dynamics, breadth-first search (\texttt{BFS}) from graph algorithms, Needleman-Wunsch (\texttt{NW}) from bioinformatics, Gaussian Elimination (\texttt{GE}) from linear algebra, stream cluster (\texttt{SC}) from data mining, and particle filter (\texttt{PF}) from the medical imaging domain. Unlike \texttt{$\mu$Kern}, these benchmarks have multiple kernels, and when a particular iteration of these kernels is running on GPU, CPU could be preparing data for the next kernel launch. Therefore, they are also called as \emph{heterogeneous} benchmarks. Among the selected heterogeneous benchmarks, \texttt{BFS} and \texttt{PF} benchmarks have high branch-divergences, so they perform better on CPU than GPU. Further, when run on GPU with CPU as host, \texttt{CFD}, \texttt{GE}, and \texttt{PF} have low CPU-load, defined as the proportion of total runtime spent on the CPU device, while others spend a large portion of their total runtime on the CPU device.

Further, we use single-threaded benchmarks from the SPEC's CPU2006 benchmarks-suite~\cite{Spradling_can2007} to contrast the thermal and power implications of CPU-only workloads against the OpenCL workloads. When a single-threaded SPEC benchmark is launched on CPU, it uses only one out of the four CPU cores. On the other hand, when an OpenCL kernel is launched on CPU cores, it uses all the available cores of the CPU device.  Hence, the single-threaded benchmarks are used to highlight the impact of core-level scheduling decisions on the CPU cores.

\vspace{-0.05in}
\section{Results}
\label{sec:results}

In this section, we present multiple implications of integrated CPU-GPU processors on thermal and power management techniques through experiments on a real processor. In subsection~\ref{sec:interplay1} and~\ref{sec:interplay2}, we demonstrate the intertwined behavior of DVFS and scheduling for OpenCL workloads on a CPU-GPU processor. Subsection~\ref{sec:asymPD} highlights the implications of asymmetric power density of two devices on thermal and power management solutions for these processors. Finally, in subsection~\ref{sec:lkgAware}, we show the importance of CPU side scheduling decisions of OpenCL workloads launched on GPU with one of the CPU cores as the host core, which executes the sequential part of the workload.

 \begin{figure*}[t!]
\begin{center}
\scalebox{0.6}{
\includegraphics[width=1.6\textwidth]{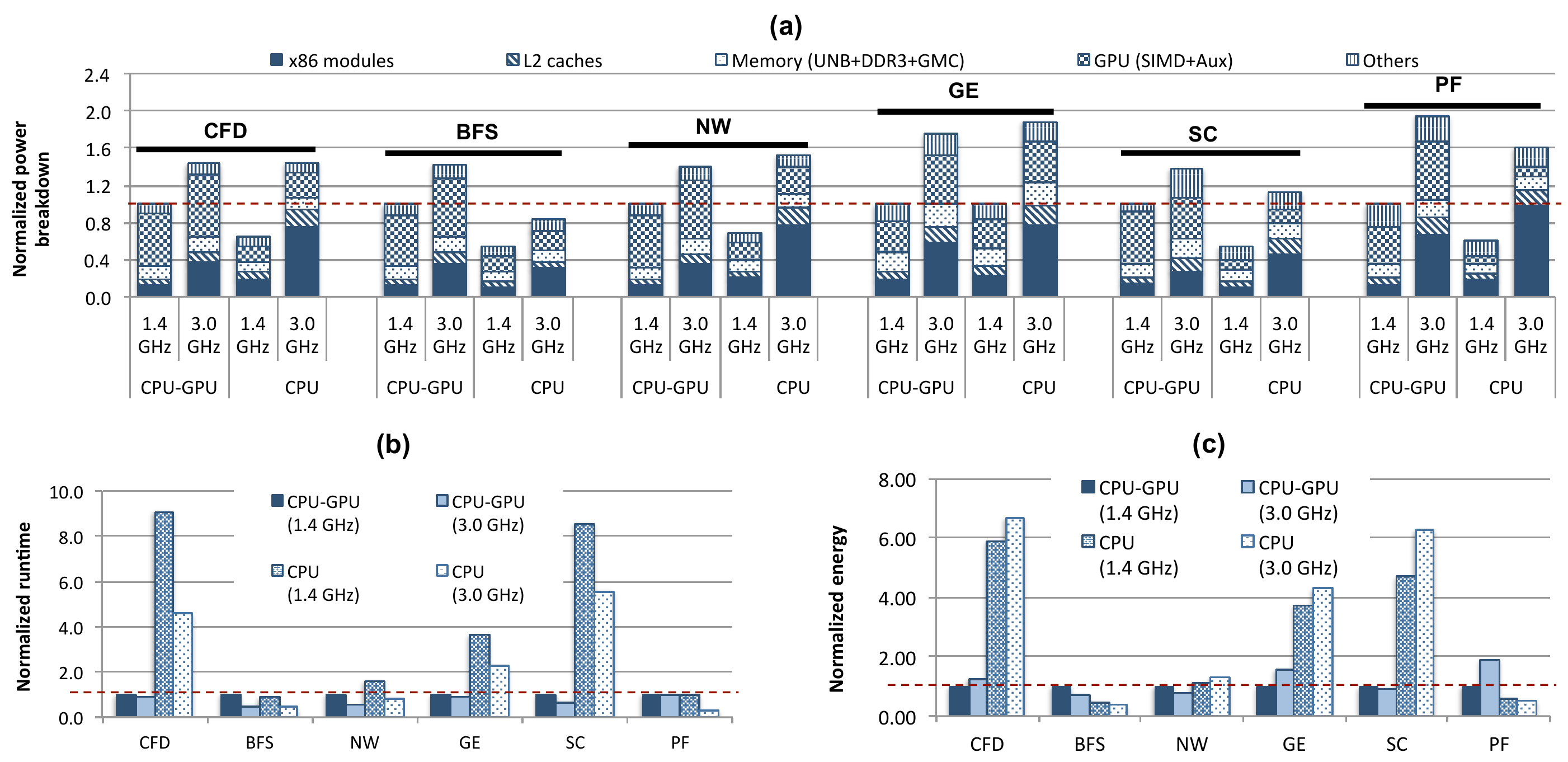}}
\vspace{-0.1in}
\caption{(a) Normalized power breakdown, (b) runtime, and (c) energy for 6 heterogeneous OpenCL benchmarks executed on CPU-GPU and CPU devices at two different CPU DVFS settings (normalization with respect to "CPU-GPU at 1.4 GHz" cases).}
\vspace{-0.25in}
\label{fig:temp_and_pwr_opencl}
\end{center}
\end{figure*}

\vspace{-0.1in}
\subsection {Interplay of Scheduling and DVFS}
\label{sec:interplay1}
To demonstrate the interactions between DVFS and scheduling on CPU-GPU processors, we analyze the thermal, power and performance characteristics of a heterogeneous OpenCL workload (\texttt{CFD}) launched on the AMD APU. 
Figure~\ref{fig:cfd_TPmaps} shows thermal maps and their corresponding power-breakdowns for two scheduling cases (CPU-GPU and CPU) and two DVFS settings (1.4 and 3.0~GHz for the CPU cores) for the \texttt{CFD} benchmark. The power breakdowns for different cases are shown in the pie-charts. As expected, the power in x86 modules is higher when the kernel is launched to CPU, while the power in GPU units is higher when it is launched to the GPU device. The figure also shows the peak temperature, total power and runtime for each case. We notice that the power, performance, and thermal profiles are different in different cases. In particular, among the four options, two DVFS settings and two scheduling choices, shown in Figure~\ref{fig:cfd_TPmaps}, the total power and the peak temperature for the OpenCL \texttt{CFD} benchmark vary up to 23.4 W and 40.5~$\celsius$, while the performance varies by a factor of 10.5$\times$. 

Further, DVFS and scheduling, collectively, have strong effect on the location of thermal hotspots. For example, when the \texttt{CFD} kernels are launched on the CPU, as expected, CPU DVFS has negligible impact on the location of thermal hotspots on the die (column 3 and 4 in Figure~\ref{fig:cfd_TPmaps}). However, when the kernels are launched on GPU, the sequential part of the workload still runs on one of the CPU cores. So, in some case, for example in the thermal map shown in column 2 of Figure~\ref{fig:cfd_TPmaps}, the thermal hotspot could be located in the x86 module even though the parallel kernels are running on GPU. This is because the maximum operating frequency of CPU cores is higher than GPU compute unites due to the CPU having deeper pipelines and smaller register files than the GPU. Also, power has a super linear relationship with the operating frequency/voltage ($\propto fV^2$), therefore, CPU cores typically have higher power density than GPU at higher frequency.  Hence, as shown in the thermal maps in Figure~\ref{fig:cfd_TPmaps}, the location of thermal hotspot  for the application-based scheduling on a CPU-GPU processor is dependent on both CPU DVFS and scheduling choices.

The strong intertwined behavior of DVFS and scheduling on performance and power does not exist in traditional multi-core processors. For example, as was shown earlier in Figure~\ref{fig:os_based_vs_app_based}, when the  \texttt{hmmer} (SPEC) benchmark is launched on different CPU cores although the location of thermal hot spot shifts (which has ramifications on thermal management), the performance, peak temperature and total power does not change significantly because all four CPU cores have the same micro-architecture. The slight differences in the total power (28.8~W versus 31.7~W) and die temperature (78~$\celsius$ versus 80.2~$\celsius$) are because of differences in leakage power arising from differences in relative proximity of these cores to the GPU units. We observed similar behavior on the other representative SPEC CPU benchmarks, namely, \texttt{omnetpp} (memory-bound integer-point), \texttt{soplex} (memory-bound floating-point), and \texttt{gamess} (compute-bound floating-point). For brevity, we showed  results for only the \texttt{hmmer} benchmark (compute-bound floating-point) in Figure~\ref{fig:os_based_vs_app_based}.

In summary, the power management for regular multi-core CPUs is simpler than for heterogeneous processors, because DVFS and scheduling can be largely considered independently in CPUs. In CPUs, the behavior of DVFS and scheduling could be intertwined from a thermal perspective with weak interactions on power due to thermal coupling on the die. However, DVFS and scheduling techniques have greater impact on performance and thermal/power profiles of CPU-GPU processors for OpenCL workloads, which can be fluidly mapped to the CPU or GPU. We summarize the discussion as the following implication.

\vspace{0.05in}
\noindent \emph{\bf Implication 1:} \emph{DVFS and scheduling must be considered simultaneously for the best runtime, power, and thermal profiles on CPU-GPU processors.}

\subsection  {Workload-Dependent Scheduling and DVFS Choices}
\label{sec:interplay2}
Different OpenCL workloads have different characteristics such as branch divergences behavior and the proportions of work distributed between CPU and GPU devices. Therefore, the optimal scheduling and DVFS choice for performance and power/temperature varies across different workloads. Below, we provide the optimal scheduling and DVFS choices for the selected heterogeneous workloads.

\vspace{0.05in}
\noindent {\bf Power and Temperature Minimization.}
In Figure~\ref{fig:temp_and_pwr_opencl}.a, we show the breakdown of the total power for the selected heterogeneous OpenCL benchmarks under different scheduling and DVFS conditions. The power values are normalized with respect to the total power in ``1.4 GHz CPU-GPU" case for each benchmark. As expected, we notice that for all benchmarks the average total power is the lowest when they are launched on CPU at 1.4 GHz. Similarly, although not shown in the figure, for all benchmarks the peak die temperature is the lowest when they are launched on CPU at 1.4 GHz. This is expected because for the ``1.4 GHz, CPU" case, CPU frequency is the lowest and GPU is idle; so, both CPU and GPU dissipate the least amount of power. In all other cases, either CPU will dissipate higher power or both CPU and GPU will dissipate power. 

\begin{table}[t!]\scriptsize
  \centering
  \caption{Optimal DVFS and scheduling choices to minimize power, runtime, and energy for the selected heterogeneous OpenCL workloads}
    \begin{tabular}{|cccc|}
    \hline
    \textbf{Workload} & \textbf{Minimum} & \textbf{Minimum} & \textbf{Minimum}\\
    \textbf{Name} & \textbf{Power/Temp} & \textbf{Runtime} & \textbf{Energy}\\    
    \hline    
    \hline
    CFD   & 1.4 GHz, CPU  & 3.0 GHz, CPU-GPU & 1.4 GHz, CPU-GPU \\
    \hline
    BFS  & 1.4 GHz, CPU   & 3.0 GHz, CPU  & 3.0 GHz, CPU \\
    \hline
    NW  & 1.4 GHz, CPU     & 3.0 GHz, CPU-GPU & 3.0 GHz, CPU-GPU \\
    \hline
    GE    & 1.4 GHz, CPU  & 3.0 GHz, CPU-GPU  & 1.4 GHz, CPU-GPU \\
        \hline
    SC    & 1.4 GHz, CPU  & 3.0 GHz, CPU-GPU  & 3.0 GHz, CPU-GPU \\
    \hline
    PF  & 1.4 GHz, CPU    & 3.0 GHz, CPU & 3.0 GHz, CPU \\
    \hline
    \end{tabular}%
  \label{tab:summary_table}%
  \vspace{-0.05in}
\end{table}%

Further, we notice that the irregular benchmarks [(e.g., \texttt{BFS}) with better power efficiency on CPU] could dissipate unnecessarily high power if run on CPU-GPU.
This is also because the current GPUs do not have fine grained (SIMD or CU-level) power gating. So, when a workload with irregular branches is launched on GPU, only a portion of SIMDs would be doing the useful work and others would be idle, dissipating unnecessary leakage power. The recent ideas related to fine-grained power gating to reduce leakage power in GPUs would be quite useful in such cases~\cite{DevISVLSI16, AroraICMMCS14, XuPACT14, DevHPC16, AroraHPCA15}. 
We summarize these observations as the following implication.

\vspace{0.05in}
\noindent \emph{\bf Implication 2:} \emph{Running workloads on CPU device at the lowest DVFS setting provides minimum power and peak temperature because power gating GPU is more power efficient than keeping both CPU and GPU active at low CPU DVFS.}

\vspace{0.05in}
\noindent {\bf Runtime and Energy Minimization.}
Figure~\ref{fig:temp_and_pwr_opencl}.b and~\ref{fig:temp_and_pwr_opencl}.c illustrate the performance and energy results of the selected heterogeneous OpenCL workloads at different scheduling and DVFS settings. The optimal DVFS and scheduling choices for minimizing runtime, energy, along with power, are summarized in Table~\ref{tab:summary_table}. 
Among them, the energy or runtime results are more interesting. We notice that the optimal scheduling for minimizing energy and runtime are typically the same, but the optimal DVFS settings for minimizing energy and runtime could be different. In other words, we make following two observations from the results shown in Table~\ref{tab:summary_table}. 

\begin{enumerate}
\item \emph{If a particular scheduling choice minimizes runtime, it also minimizes the energy.} From Table~\ref{tab:summary_table}, we observe that running \texttt{BFS} and \texttt{PF} on a CPU leads to both optimal energy and runtime. However, \texttt{CFD}, \texttt{NW}, \texttt{GE}, and \texttt{SC} lead to lower energy and runtime when run on a GPU with a CPU as the host device. This behavior is observed because \texttt{BFS} and \texttt{PF} have high control-divergences. Therefore, they are more suited for the CPU architecture. All other benchmarks have high parallelism; thus, they are suited for GPU. 
\item \emph{The energy of CPU-GPU benchmarks with low CPU-load could be minimized by reducing the CPU frequency.} We observe that the energy of \texttt{CFD} and \texttt{GE} is the lowest at low CPU frequency (1.4 GHz). This is because \texttt{CFD} and \texttt{GE} have low CPU-load, measured by the relative portion of the work executed on CPU when the workload is launched on GPU. So, the performance improvement from increasing the CPU frequency does not compensate for the corresponding increase in power for these benchmarks. On the other hand, \texttt{NW} and \texttt{SC} have the lowest energy at high CPU frequency (3.0 GHz). This is because \texttt{NW} and \texttt{SC} have high CPU-load, so, increasing the CPU frequency improves their performance significantly. 
\end{enumerate}

We summarize these results through following implication.

\vspace{0.05in}
\noindent \emph{\bf Implication 3:} \emph{The optimal DVFS and scheduling choices for minimizing runtime and energy on a CPU-GPU processor are functions of workload characteristics.}

\subsection {Asymmetric Power Density of CPU-GPU Processor}
\label{sec:asymPD}
\begin{figure}[t!]
\begin{center}
\scalebox{1}{
\includegraphics[width=0.4\textwidth]{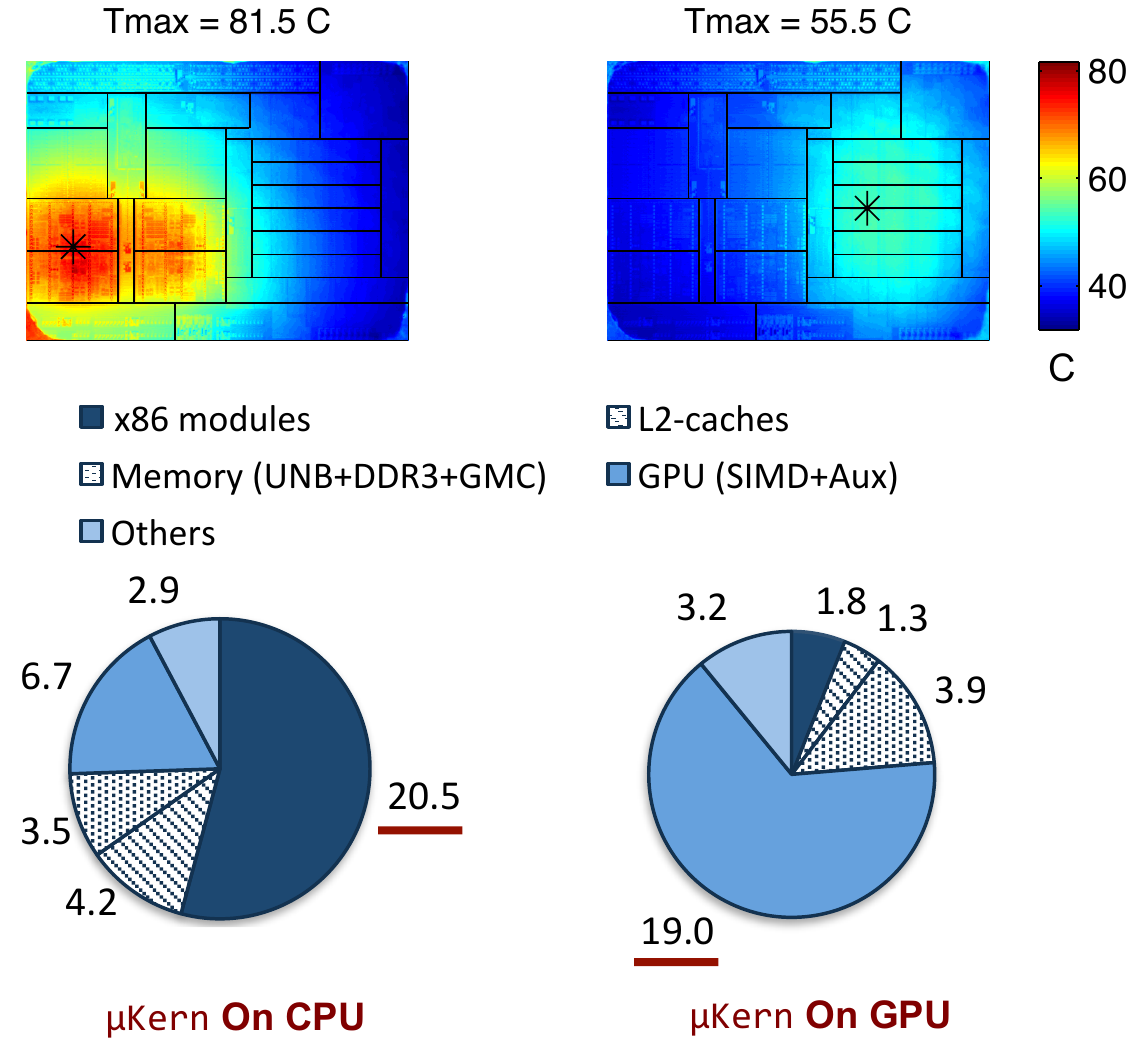}}
\vspace{-0.05in}
\caption{Thermal and power maps demonstrating asymmetric power density of CPU and GPU devices. \texttt{$\mu$Kern} is launched on CPU and GPU devices. For the comparable power on CPU (20.5 W) and GPU (19 W), the peak temperature on CPU is about 26~$\celsius$ higher than on GPU.}
\vspace{-0.25in}
\label{fig:uKern_tpmaps}
\end{center}
\end{figure}

Typically, power dissipation in GPU and CPU devices for the same OpenCL kernel is different due to differences in their architectures and operating frequencies. Further, for the studied heterogeneous processor, GPU occupies larger die-area than the CPU, and therefore, it has lower power density than CPU for the same total power. In this section, we confirm the asymmetric power density of two device experimentally. Although it is difficult to make a circuit block to dissipate a certain amount of power in a real processor, the homogeneous \texttt{$\mu${Kern}}, which keeps only one device active at a time, is used to analyze the power densities of the two devices. Figure~\ref{fig:uKern_tpmaps} shows the thermal and power maps of the die when we launch the \texttt{$\mu${Kern}} on CPU and GPU devices at fixed DVFS (3~GHz). From the pie-charts, we observe that the power consumption of CPU in column-1 (20.5~W) is comparable to the power consumption of GPU in column-2 (19~W). However, from the thermal maps, we notice that the peak temperature in two cases are significantly different; in particular, the peak temperature of CPU is higher than that of GPU by 26~$\celsius$. We computed the power density of CPU and GPU in two cases and found that the power density of CPU is $2.2\times$ higher than that of the GPU. Therefore, even for the comparable amount of power, CPU has higher peak temperature than the GPU.

 \begin{figure*}[t!]
\begin{center}
\scalebox{1}{
\includegraphics[width=1.00\textwidth]{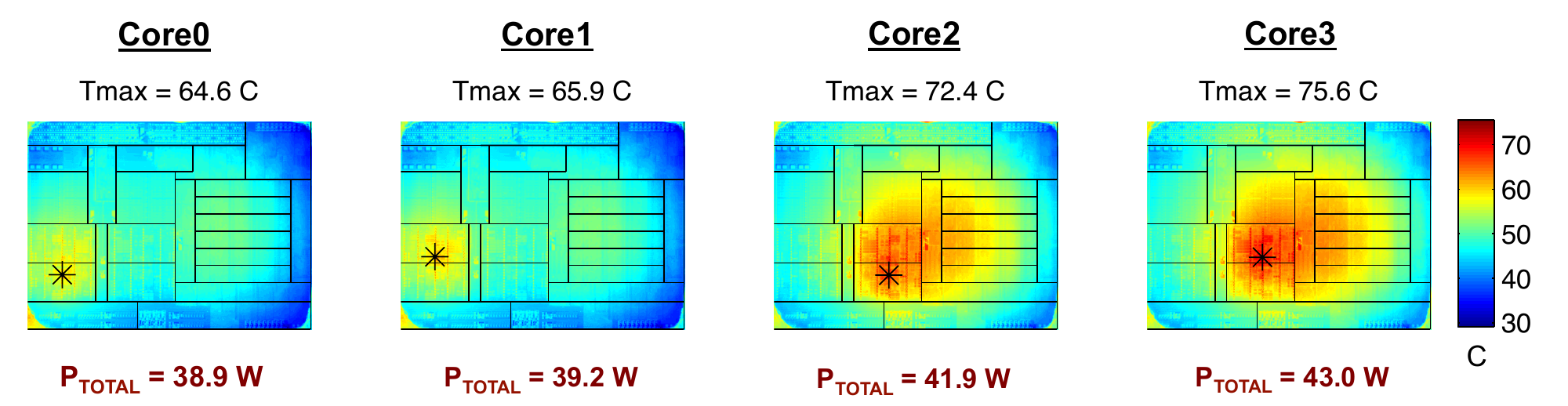}}
\vspace{-0.1in}
\caption{Impact of CPU core-affinity when a benchmark (\texttt{SC}) is launched on GPU from different CPU cores at fixed DVFS setting.}
\vspace{-0.25in}
\label{fig:sc_tmaps_c0123}
\end{center}
\end{figure*}

Further, due to higher power density of CPU than GPU, it is possible that the thermal hotspot could be located on CPU even though the OpenCL kernels are launched on GPU. This is because when a kernel is launched on GPU, CPU acts as its host device; so, the CPU could also be active to prepare the work for the next iteration of kernel launch. This could be seen from the thermal map shown earlier in the column-2 of Figure~\ref{fig:cfd_TPmaps}. Notice at 3~GHz DVFS, the peak temperature is located on CPU blocks even though the kernels are launched on GPU. More importantly, we observe that, at higher DVFS state, the likelihood of CPU reaching the thermal limit first is higher than that of GPU.  Although at low DVFS (e.g., 1$^{st}$ column of Figure~\ref{fig:cfd_TPmaps}), the hotspot may be located on GPU blocks, but the peak temperature of GPU in that case is lower than the thermal limits. The higher temperature of GPU in this case would lead to higher leakage power, but the peak temperature of GPU will still be below the thermal limit.

The asymmetric power density and its effect on thermal profiles of the CPU-GPU processor, as discussed above, could have multiple implications on the thermal and power management  of the system. Few of them are as follows.

\vspace{0.05in}
\noindent \emph{\bf Implication 4:} \emph{Due to lower peak temperature in GPU, it could have fewer number of thermal sensors per unit area than the CPU.}

\vspace{0.05in}
\noindent \emph{\bf Implication 5:} \emph{The extra thermal slack available on the GPU could be used to improve its performance through frequency-boosting, provided it meets all architectural timings, like register file access time.}

\vspace{0.05in}
\noindent \emph{\bf Implication 6:} \emph{One could design a localized cooling solution (e.g., thermo electric cooler based) for separate and efficient cooling of CPU and GPU devices on such processors.}

\vspace{-0.15in}
\subsection {Leakage Power-aware CPU-side Scheduling}
\label{sec:lkgAware}
In this subsection, we demonstrate the importance of scheduling the sequential part of an OpenCL CPU-GPU workload on an appropriate core of the CPU. Typically, a CPU-GPU processor has multiple cores on the CPU side. So, when a workload is launched on GPU with CPU as the host device, it could be launched from any of the available CPU cores. Since the cores in x86 module-2 (M2) are in close proximity to GPU than the cores in x86  module-1, the thermal coupling, and therefore, the leakage power is different in each case.   

To understand the differences in thermal and power profiles, we launched a heterogeneous benchmark (\texttt{SC}) on GPU from 4 different cores of the CPU. Figure~\ref{fig:sc_tmaps_c0123} shows the thermal maps of the die in all 4 cases. We observe that the thermal and power profiles of the chip is indeed different for different core-affinities. Specifically, the total power when the workload is launched from Core0, 1, 2, and 3 are 38.9~W, 39.2~W, 41.9~W, and 43~W, respectively; while, the corresponding peak die temperature values are 64.6~$\celsius$, 65.9~$\celsius$, 72.4~$\celsius$, and 75.6~$\celsius$, respectively. Hence, we notice that the total power and peak temperature of the die is higher (by 4~W and 11~$\celsius$, respectively) when the benchmark is launched from Core3 than when it is launched from Core0. This happens because Core3 is in close proximity to GPU; so, there is stronger thermal coupling between Core3 and GPU than between Core0 and GPU. The strong thermal coupling leads to higher temperature and leakage power in both CPU and GPU. So, it is important to launch the kernels from an appropriate CPU core. We encapsulate this observation in the following implication.

\vspace{0.05in}
\noindent \emph{\bf Implication 7:} \emph{The OS or the CPU-side scheduler should use the floorplan information of the processor to launch a workload on GPU from an appropriate CPU-core to reduce both, the peak temperature and the leakage power of the chip.}

\vspace{-0.05in}
\section{Conclusions and Future Directions}
\label{sec:conclusions}

CPU-GPU processors are becoming mainstream due to their versatility in terms of performance and power tradeoffs. In this paper, we showed that the integration of two architecturally different devices, along with the OpenCL programming paradigm, create new challenges and opportunities to achieve the optimal performance and power efficiency for such processors. 
We highlighted multiple implications of CPU-GPU processors on their thermal and power management techniques with the help of detailed thermal and power breakdown.
We demonstrated that in comparison to traditional multi-core CPU benchmarks, the OpenCL paradigm enables more diverse thermal and power management decisions, leading to greater performance and thermal/power tradeoffs on a heterogeneous processor. 
We elucidated the impact of workloads and power management decisions on the locations of hot spots for effective thermal management. 

We will extend our work by studying the impact of adaptively changing the operating frequency and the number of active compute-units in the GPU based on the workload characteristics. 
Further, we would build models to predict optimal scheduling and DVFS choices for OpenCL workloads, and design a comprehensive thermal and power management scheme for heterogenous CPU-GPU processors leveraging the key findings of this work.


\ifCLASSOPTIONcompsoc
\else
\fi


\ifCLASSOPTIONcaptionsoff
  \newpage
\fi

\bibliographystyle{IEEEtran}
\vspace{-0.1in}
\bibliography{references}

\end{document}